# Mobility overestimation in MoS$_2$ transistors due to invasive voltage probes

*Peng Wu[1]\**

[1]Research Laboratory of Electronics, Massachusetts Institute of Technology, Cambridge, MA 02139, USA

*E-mail: pengw@mit.edu

Improving carrier mobilities of two-dimensional (2D) semiconductors is highly sought after. Recently, Ng. et al.[1] reported rippled molybdenum disulfide (MoS$_2$) transistors on bulged silicon nitride (SiN$_x$) substrates that exhibit high electron mobilities up to ~900 cm$^2$V$^{-1}$s$^{-1}$. The high mobility values were attributed to the suppression of electron-phonon scattering by the lattice distortion in the rippled MoS$_2$ channel. While the results are compelling, this Matters Arising shows that the mobility values in ref. [1] are likely to be overestimated due to invasive voltage probes in the four-probe measurement setup, which causes a positive threshold voltage shift near the voltage probes and an artificial overestimation of apparent field-effect mobility.

Previous studies[2,3] has shown that the gate-voltage dependent contact resistance in transition metal dichalcogenide (TMD) transistors could lead to an overestimation of mobility. In particular, ref. [3] shows that the threshold voltage $V_{th}$ near the metal contacts shows a positive shift compared with the $V_{th}$ in the channel region. The shift is caused by depletion of carriers near the metal contacts due to Fermi level pinning at the metal-TMD interfaces, and thus a larger gate voltage is required to populate carriers in the region (i.e., a larger $V_{th}$). Although a four-probe measurement setup is adopted in ref. [1] to exclude the impact of contact resistance, due to the invasiveness[4–6] of the voltage probes (Supplementary Fig. 14a in ref. [1]), the impact of metal contacts is not completely eliminated and the regions near the voltage probes are still expected to exhibit a positive $V_{th}$ shift (Fig. 1a).

To prove the existence of positive $V_{th}$ shift in the MoS$_2$ field-effect transistor (FET), Fig. 1b shows the carrier concentrations versus gate overdrive voltage $V_G$ - $V_{th}$ in Hall and FET measurements from Fig. 3f in ref. [1]. One can clearly see that the $V_{th}$ from FET measurement exhibit a ~25 V positive shift compared with the $V_{th}$ from Hall measurement. This is because in a two-probe FET measurement setup (or in a four-probe setup with invasive voltage probes), $V_{th}$ is determined by the region with the largest $V_{th}$, i.e., the regions near the metal contacts (or voltage probes), while in the Hall measurement, the carrier concentration is extracted from Hall resistance $R_{xy}$ versus magnetic field $B$, which measures $V_{th}$ in the channel, since the Hall voltage $V_{xy}$ is perpendicular to the current flow and is not affected by the more resistive regions near the voltage probes (inset of Fig. 1b). Similar $V_{th}$ discrepancies in Hall and four-probe measurements have also been observed in Supplementary Fig. 11 of ref. [7], a previous paper that also reported mobility enhancement in $MoS_2$ on crested $SiN_x$ substrates.

Next, a device model that captures the $V_{th}$ shift effect is developed. As shown in Fig. 1a, the total channel resistance is modelled as the series resistance of the two regions close to voltage probes (with a threshold voltage of $V_{th1}$) and the unaffected region in the channel (with a threshold voltage of $V_{th2}$). Since four-probe measurement is adopted, no contact resistance is considered. Therefore, the electrical conductivity of the channel $\sigma_{ch}$ is given by:

$$\sigma_{ch} = [\phi \sigma_1^{-1} + (1-\phi)\sigma_2^{-1}]^{-1} \tag{1}$$

in which $\phi$ is the proportion of the combined length of two regions close to voltage probes (blue regions in Fig. 1a) to the total channel length $L$, $\sigma_1$ and $\sigma_2$ are the conductivities of the regions close to voltage probes (blue regions in Fig. 1a) and the channel region (red region in Fig. 1a), respectively, given by[8]:

$$\sigma_i = \frac{L_i}{W t_{body}} G_i = \frac{\mu_{model}}{t_{body}} C_{ox} \frac{n k_B T}{q} \ln\left[1 + \exp\frac{q(V_G - V_{th,i})}{n k_B T}\right], i = 1, 2 \tag{2}$$

in which $n = 1 + q^2 D_{it}/C_{ox}$ is the band movement factor, $D_{it}$ is interface trap density. A constant mobility $\mu_{model}$ is assumed in the model. The parameters for the model are listed in Table 1.

**Table 1 | Parameters for the model**

| |
|---|
| $\mu_{model}$ = 140 cm$^2$V$^{-1}$s$^{-1}$, $C_{ox}$ = 22 nF/cm$^2$ (given in ref. [1]), $t_{body}$ = 1.4 nm (bilayer MoS$_2$) |
| $\phi$ = 0.1, $V_{th1}$ = 1.5 V, $V_{th2}$ = -30 V, $D_{it}$ = 3.5 ×10$^{12}$ cm$^{-2}$eV$^{-1}$, $T$ = 300 K |

Fig. 1c shows a comparison of modelled and experimental electrical conductivity $\sigma_{ch}$ of $MoS_2$ versus gate voltage $V_G$, where the experimental data are extracted from Fig. 3b in ref. [1]. The conductivities of component segments, $\sigma_1$ and $\sigma_2$, are also shown for comparison. One can see that the model shows good agreement with the experimental data. The curvatures of the $\sigma_{ch} - V_G$ curve at different gate voltages are well reproduced, which proves that the model captures the essential mechanism of device operation. In addition, the $\sigma_{ch} - V_G$ curve (purple curve) exhibits a much steeper slope than the $\sigma_1 - V_G$ curve (blue curve) or the $\sigma_2 - V_G$ curve (red curve). This indicates that an overestimation of mobility is expected when extracting the apparent field-effect mobility $\mu_{FE,app}$ from the slope of the $\sigma_{ch} - V_G$ curve using:

$$\mu_{FE,app} = \frac{\partial I_D}{\partial V_G} \frac{1}{V_{DS}} \frac{1}{C_{ox}} \frac{L}{W} = \frac{\partial \sigma_{ch}}{\partial V_G} \frac{1}{V_{DS}} \frac{1}{C_{ox}} t_{body} \qquad (3)$$

Fig. 1d shows a comparison of modelled and experimental $\mu_{FE,app}$ of MoS$_2$ versus $V_G$, where the experimental data are extracted from Fig. 3b in ref. [1] and the modelling results are calculated using eq. (3). The modelled $\mu_{FE,app}$ shows good agreement with experimental data across different $V_G$ ranges, with the only exception of from 20 V to 30 V. Note that $\mu_{FE,app}$ exceeds the actual mobility used in the model, $\mu_{model}$ = 140 cm$^2$V$^{-1}$s$^{-1}$, which highlights the mobility overestimation in the measurement. Moreover, even with a constant $\mu_{model}$, the interplay of $\sigma_1$ and $\sigma_2$ at different $V_G$ in the model can readily explain the change of $\mu_{FE,app}$ over a broad range of $V_G$ (except for from 20 V to 30 V), without involving a complicated transitioning between different dominant scattering mechanisms at different $V_G$ range that is proposed in ref. [1], thus providing a simpler and more natural explanation[9] to experimental observations. Note that similar trends of $\mu_{FE,app}$ change with $V_G$ were also observed in ref. [7], which also indicates a possible mobility overestimation.

Finally, to provide readers a guidance on mobility measurement using four-probe measurement, different configurations of four-probe measurement are compared in Fig. 2. Previous discussion in this study mainly focuses on the first configuration with invasive voltage probes, in which the voltage probes extend over the entire channel width direction and cause the largest disturbance to the current flow and voltage distribution in the channel. Therefore, the first configuration is not recommended for four-probe measurement, although it can be used in transfer length method[10] (TLM) to more reliably exclude the impact of metal contacts.

The second configuration with partially invasive voltage probes is adopted in part of the results in ref. [1] (Fig 3a and Supplementary Fig. 21 in ref. [1]). However, the voltage probes still disturb the current flow and voltage distribution in the channel, and the measured longitudinal voltage $V_{xx}$ may pick up signals from the voltage drop in the more resistive depletion regions near the voltage probes, and thus the measurement is still prone to mobility overestimation as discussed before. That being said, this configuration may still be adopted if one tries to minimize the overlap between the voltage probes and the channel[3,11] and reduce the disturbance from the voltage probes as much as possible.

The third configuration with non-invasive voltage probes is most recommended for four-probe measurement. In this configuration, the impact of voltage probes to the current flow and voltage distribution in the channel is minimized, although the device fabrication is more complicated as it involves etching the channel into a Hall bar geometry[12].

In conclusion, the mobility values reported in ref. [1] (and in a previous paper[7]) are likely to be overestimated due to the impact of invasive voltage probes in the four-probe measurement. A device model that captures the impact of the metal contacts is developed, which achieves good agreement with the experimental data while providing a simple and convincing explanation for the apparent field-effect mobility ($\mu_{\text{FE,app}}$) change with gate voltage.

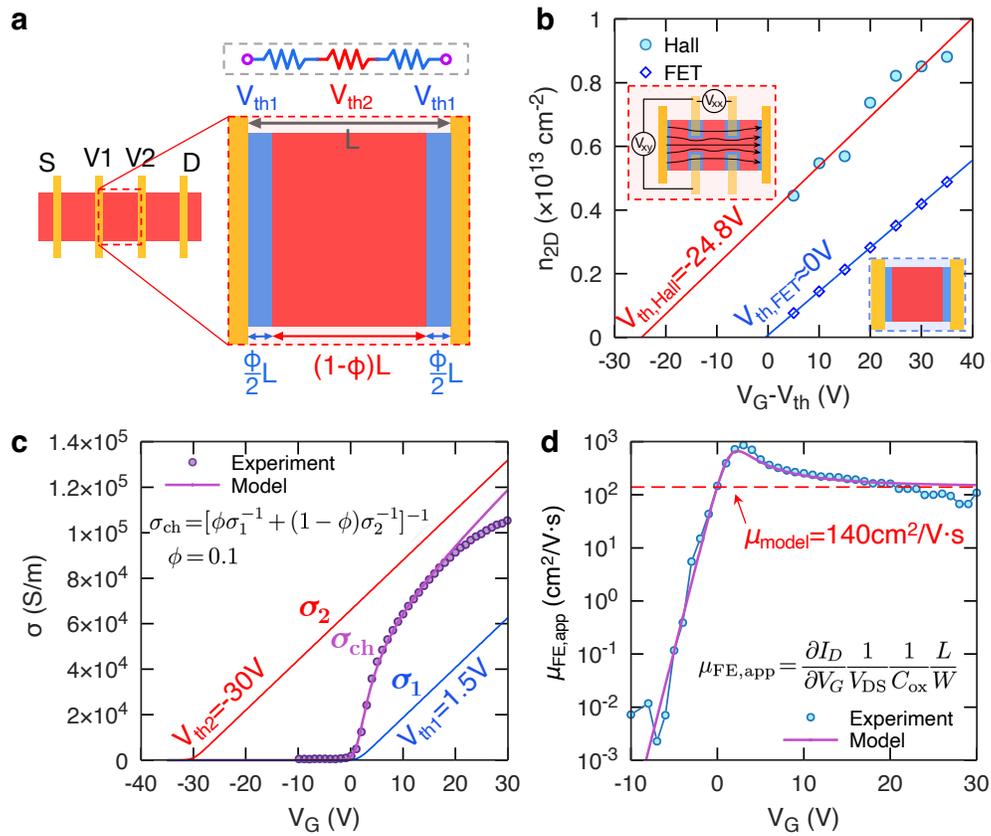

**Fig. 1 | Mobility overestimation due to invasive voltage probes and threshold voltage shift. a,** Schematic of four-probe measurement setup and illustration of threshold voltage shift caused by the invasive voltage probes. **b,** Carrier concentrations versus gate overdrive voltage in Hall and FET measurements. Experimental data are extracted from Fig. 3f in ref. [1]. **c,** Modelled and experimental electrical conductivity of $MoS_2$. Inset shows the threshold voltage distribution in the channel used in the model. Experimental data are extracted from Fig. 3b in ref. [1]. **d,** Modelled and experimental apparent mobility of $MoS_2$. Experimental data are extracted from Fig. 3b in ref. [1].

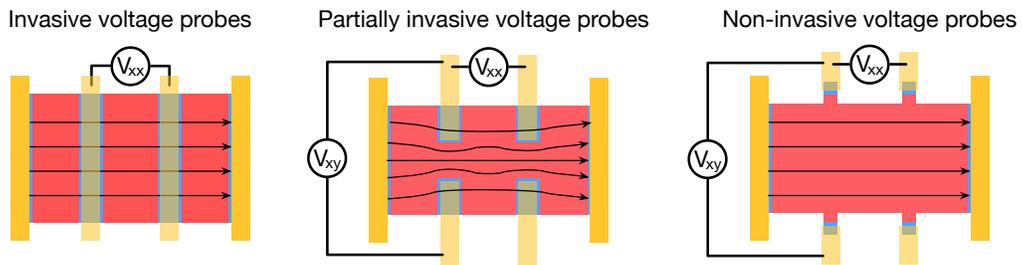

**Fig. 2 | Comparison of different configurations of four-probe measurement.**